\documentstyle[12pt,aasms4]{article}
\begin{document}

\title{The Infrared and Optical Variability of OJ287}

\author{J.H. Fan and G. Adam}
\affil{  CRAL Observatoire de Lyon, 9, Avenue Charles André, 69 563 St-Genis-Laval Cedex, France}

\author{G.Z. Xie}
\affil{ Yunnan Observatory, Chinese Academy of Sciences, Kunming 
650011, China} 

\author{S.L. Cao}
\affil{ Department of Astronomy, Beijing Normal 
University, Beijing, China}

\author{R.G. Lin}
\affil {Center for Astrophysics, Guangzhou Normal University, Guangzhou 510400, China} 

\author{Y.P. Qin}
\affil{ Yunnan Observatory, Chinese Academy of Sciences, 
 Kunming 650011, China } 

\author{Y. Copin}
\affil{ CRAL Observatoire de Lyon, 9, Avenue Charles André,
 69 563 Saint-Genis-Laval Cedex, France}

\author{J.M. Bai, X. Zhang, and K.H. Li}
\affil{Yunnan Observatory, Chinese Academy of Sciences, 
 Kunming 650011, China}

\begin{abstract}
 In this paper, the long-term historical optical (UBVRI)  and near-infrared (JHK) data 
 are presented with some new observations in the optical (February 1994-January 1995) and 
 near-infrared (November 1995) bands included for BL Lac object OJ287. The new optical 
 data in V-band are in agreement with the results reported by other authors (Sillanpaa et al. 
 1996a; Arimoto et al. 1997), a close correlation between the color index of B-V and the 
 magnitude V has been obtained from our new observations; The new infrared 
 observations  presented here indicate that the source was at a high level in the infrared
 band during the observation period; From the available literature, we have got that the largest 
 variations for UBVRIJHK bands are respectively: $\Delta U = 4^{m}.72$;
 $\Delta B = 5^{m}.93$; $\Delta V = 5^{m}.18 $
 $\Delta R = 4^{m}.45 $; $\Delta I = 4^{m}.07$;
 $\Delta J = 3^{m}.87 $; $\Delta H = 3^{m}.78$;
 $\Delta K = 3^{m}.54 $. A strong correlation is found between the optical and near-infrared
 bands when the DCF method is used, which suggests that these two bands have the same emission mechanism.
\end{abstract}
\keywords{Variability -- Correlation -- BL Lacertae Object - individual OJ 287}

\section{Introduction}
 BL Lac objects are a special subclass of active galactic nuclei (AGNs), which show some extreme 
 properties: rapid and large variability, high and variable polarization, no or only weak emission 
 lines in its classical definition.

 BL Lac objects are variable not only in the optical band, but also in radio, infrared,  X-ray, 
 and even $\gamma$-ray bands. Some BL Lac objects show strongly correlated variation between 
 radio and optical emissions with some delay (e.g. Tornikoski et al. 1994); Some BL
 Lac objects show that the spectral indices change with the brightness of the sources 
 (Bertaud et al. 1973; Brown et al. 1989; Fan 1993), generally, the spectrum flattens when 
 the source brightens, but different phenomenon has also been observed, from 3C66A for 
 instance (see De Diego et al. 1997).

 The nature of AGNs is still an open problem; the study of AGNs variability can 
 yield valuable information about their nature, and the implications for quasars modeling 
 are extremely important(Blandford, 1996).  From the telescopes in China (the optical 
 telescopes: 1-m telescope at Yunnan Observatory, the 1.56-m telescope at Shanghai Observatory, 
 and the 2.16-m telescope at Beijing Observatory; and the 1.26m infrared telescope at 
 Beijing observatory), we have monitored dozens of AGNs, including BL Lac objects, Quasars, 
 and Seyfert galaxies (Xie et al. 1987, 1988a,b, 1990, 1991, 1992, 1994;
 Fan et al. 1997, Bai et al. 1998, Xie et al. 1998).

 OJ287 (VRO 20.08.1) was discovered in radio observation by Dickel et al. (1967) and 
 in the optical band by Thompson et al. (1968). Spectroscopic observations of Miller et al. 
 (1978) showed a weak  $[O_{III}]$ spectral feature with a redshift of 0.306, which was
 confirmed by later observation of Sitko \& Junkkarinen (1985) and Stickel et al. (1989). 
 Early observations indicated that it was variable in the radio (Blake 1970) and optical 
 bands (Kinmin \& Conkin 1971). It was observed to show only continuum emission 
 (Adam et al. 1972) and high linear polarization (Kinman \& Conkin 1971, Nordsiek 1972). 
 It has been observed extensively since. 

 OJ287 is one of the very few AGNs for which continuous light curve over more than 
 one hundred years has been observed. Its observational properties from radio to X-ray 
 have been reviewed by Takalo (1994). The light curve in optical band shows the strong  
 signature of an  outburst that occurs with a period of 12 years. Models including 
 binary black hole can explain these variations (Sillanpaa et al. 1988a). The outburst 
 predicted to occur in the fall of 1994 confirmed the 12-year  period (Sillanpaa et al. 1996b,c). 

 Besides the long-term large outburst, OJ287 has also shown some interesting fluctuations 
 of brightness on times ranging from less than 1 hour ( Visvanathan et al 1973; Veron \& 
 Veron 1975; Carrasco et al. 1985) to about a week ( Kinman et al. 1974). It is one of the
 objects in our optical monitoring program ( see Xie et al. 1994; Bai et al. 1998).

 Large infrared variations have been seen in the source. Variations of 0.3 to 0.5 
 magnitudes in the near-infrared bands over a time scale of one day have been reported 
 during its outburst by Holmes et al(1984a) and during its low state by Wolstencroft et al. (1982)
 and Impey et al. (1984).  Lorenzetti et al. (1989) reported variations of $0^{m}.3$ over 
 a time scale of 3 hours and $0^m.5$ over a day and $\langle$ J-H $\rangle$ = 0.83, 
 $\langle$ H-K $\rangle $= 0.91. As reported in the optical band ( Takalo 
 \& Sillanpaa 1989), a correlation between color index and magnitude has been found in 
 the infrared band: Gear et al. (1986a) found a correlation between the infrared spectral 
 index and the J band flux from data covering the period February 1983 to February 1986 
 after the 1983 outburst. Kidger et al. (1994) and Zhang \& Xie (1996) also found the correlation.  
 But no such significant correlation was found in the data of 1986 February through 1987 
 December (Lorenzetti et al. 1989).   In November 1995, it was detected in our observation 
 program at Xinglong station of Beijing Observatory ( see Xie et al. 1998 for detail).

 In this paper, we will present both historical and new data in the optical and infrared bands, 
 and deal with them by means of the DCF method.  The paper has been arranged as follows: 
 In section 2, we will present the observations; in section 3, the correlation analysis
 method (DCF); in section 4, we will discuss our results; and in section 5, 
 we will give a brief conclusion.

\section{Observations}
\subsection{New Observation}

 During 1994 and 1995, OJ287 was observed with the 1-m RRC telescope at Yunnan Observatory,  
 which is equipped with a direct CCD camera at the Cassergrain focus, and the 1.26-m Infrared 
 telescope at Xinglong station of Beijing Observatory. The daily averaged magnitudes 
 are presented in table 1 and 2 .  The data reduction and the detailed  magnitudes 
 will appear in separate papers altogether with other BL Lac objects observed during 
 the same period ( Bai, et al. 1998 for the optical data and Xie et al. 1998 for infrared data). 

\begin{table}
\caption[]{Daily Averaged Magnitudes of OJ 287}
\begin{tabular}{ccc}
\hline\noalign{\smallskip}
Date       &  B          &  V   \\ 
\noalign{\smallskip}
\hline\noalign{\smallskip}
94-02-07   & 16.07(0.03) &                \\ 
94-02-08   &             & 15.62(0.03)    \\
94-04-18   & 15.96(0.01) &                 \\
94-12-04   & 15.52(0.01) & 15.04(0.01)     \\ 
94-12-05   & 15.75(0.05) & 15.08(0.04)    \\ 
94-12-06   & 15.49(0.08) & 14.85(0.04)     \\ 
95-01-23   & 15.55(0.09) & 15.15(0.07)     \\ 
95-01-24   & 15.57(0.07) & 15.35(0.07)     \\ 
95-01-25   & 15.78(0.08) &  15.49(0.07)    \\ 
95-01-26   & 15.53(0.08) & 15.20(0.05)      \\ 
95-01-27   &             & 15.32(0.05)     \\ 
95-01-28   & 15.66(0.08) & 15.28(0.05)     \\ 
\noalign{\smallskip}
\hline    
\end{tabular}
\end{table}

\begin{table}
\caption[]{Daily Averaged Magnitudes of OJ 287}
\begin{tabular}{cccc}
\hline\noalign{\smallskip}
Date       &  J          &  H           & K    \\ 
\noalign{\smallskip}
\hline\noalign{\smallskip}

95-11-16   & 10.19(0.16)   & 10.45(0.17) &  9.40(0.14) \\
95-11-17   &  9.91(0.26)   &  9.78(0.38) &  9.13(0.13) \\
95-11-18   & 11.13(0.31)   & 10.83(0.40) &  9.67(0.38) \\
\noalign{\smallskip}
\hline    
\end{tabular}
\end{table}

\subsection{Historic observations}
 Since no infrared observation is available before 1971, and in order to compare 
 the optical data with the infrared data, only the observations in optical band 
 after 1971 have been compiled ( Martynov, 1971; Huruhata, 1971; Burkhead 1971; 
 Locher, 1972; Strittmatter et al. 1972; Craine \& Warner 1973; Goldsmith \& 
 Weistrop 1973;  Visvanathan 1973; Frohlich et al. 1974; Selmes et al. 1975; 
 Veron \& Veron 1975;  Kikuchi, et al. 1976; O'Dell et al. 1978a,b;   Usher 1979; 
  Hagen-Thorn et al. 1980; Pushell \& Stein 1980;  Shaefer 1980; Gaida \& Roser 
 1982; Takalo 1982; Haarala et al. 1983; Bortle 1983a,b; Sitko et al. 1983; Corso 
 et al. 1984, 1986, 1988; Holmes et al. 1984a;  Moles et al. 1984, 1985;
 Sillanpaa et al. 1985; Sitko et al. 1985; Sitko \& Junkkarinen 1985; Brindle et al. 
 1986; Cayatte et al. 1986; Miller 1987; Monella 1987; Monella \& Verdenet 1987; 
 Sillanpaa 1987; Smith et al. 1987; Xie et al. 1987, 1988a,b, 1989, 1990, 1991, 1992, 
 1994; Sillanpaa 1988a,b; Webb et al. 1988; Brown et al. 1989;Takalo 1989;  
 Mead et al. 1990; Takalo et al. 1990; Sillanpaa et al. 1991a,b;  Sitko \& Sitko 1991; 
 Takalo 1991, Takalo et al. 1992a; Valtaoja et al. 1991; Carini et al. 1992; 
 Kidger \& Takalo 1993; Okyudo 1993;  Smith et al. 1993; Fiorucci \& Tosti, 1994a,b;  
 Meusinger 1994a,b;  Vanmunster \& Cautern 1994; Kidger 1995; Arimoto et al. 1997). 
 All those data and the data presented in table 1 are used in our analysis.  
 The data in UBVRI bands are shown in Figure 1a,b,c,d.e.

 The infrared (JHK) are from  Epstein, et al. 1972; Rieke et al. 1977; O'Dell et al. 
 1977, 1978a; Pushell \& Stein 1980; Allen et al. 1982; Wolstencroft et al. 1982; 
 Impey et al. 1982, 1984; Sitko et al. 1983; Holmes et al. 1984a,b; Brindle et al. 1986;
 Gear et al. 1986a,b;  Landau et al. 1986; Roelling, et al. 1986;  Smith, et al. 1987; 
 Impey \& Neugebauer 1988; Brown et al. 1988; Lorenzetti et al. 1989;   Mead et al. 1990; 
 Takalo et al. 1992b; Bersanelli et al. 1992; Sitko \& Sitko 1991; Gear 1993; 
 Kidger et al. 1994; Litchfield et al. 1994. Some  early data  are derived from the 
 figure of Soifer \& Neugebauer (1980). The infrared (JHK) data are shown in Figure 2a,b,c.

\begin{figure} 
\epsfxsize=15cm
$$
\epsfbox{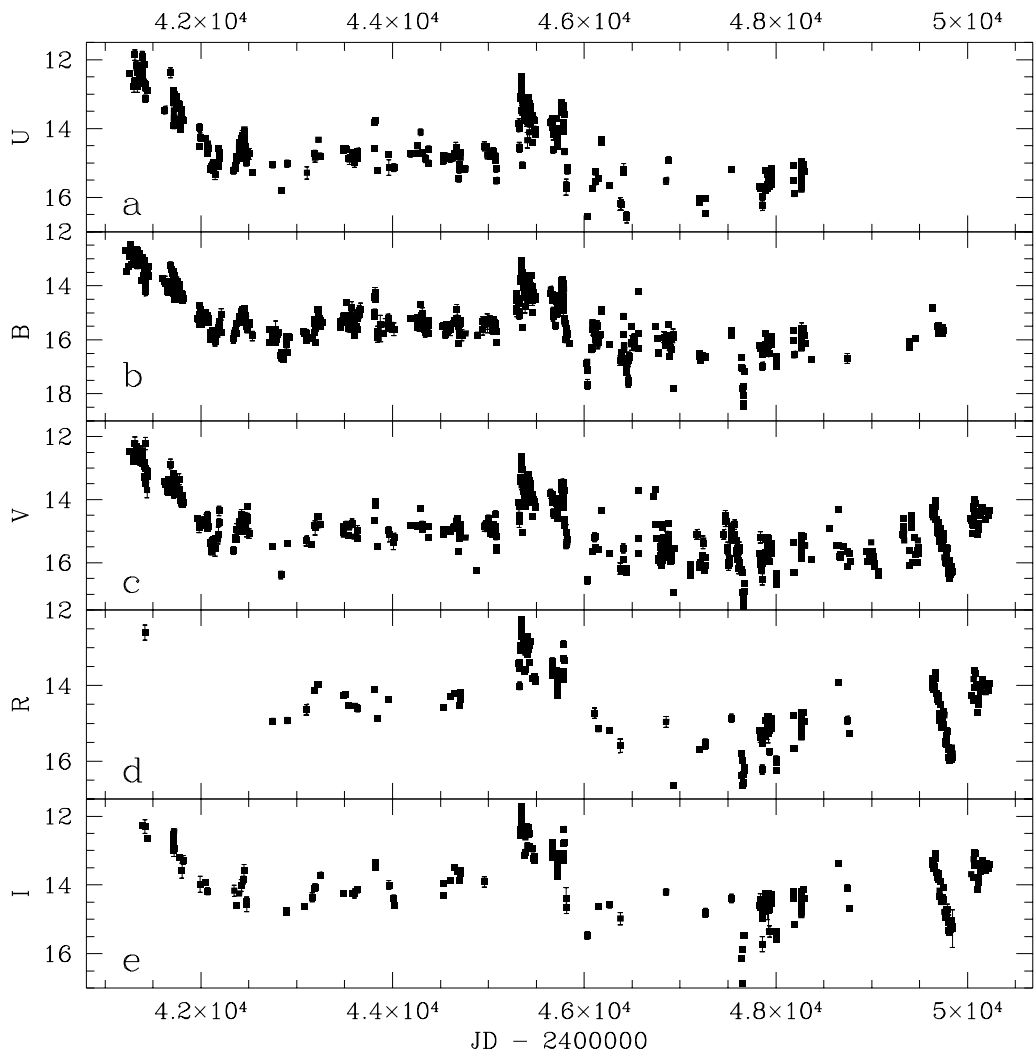}
$$

\caption{a: The long-term U light curve of OJ287 covering the period 
1971--1991, there are no new data available in the literature after 1991,
b:The long-term B light curve of OJ287  from 1971 to 1995,
c: The long-term V light curve of OJ287 from 1971 to 1996,
d: The long-term R light curve of OJ287 from 1972 to 1996,
e: The long-term I light curve of OJ 287  from 1972 to 1996 }
\end{figure}

\begin{figure}
\epsfxsize=15cm
$$
\epsfbox{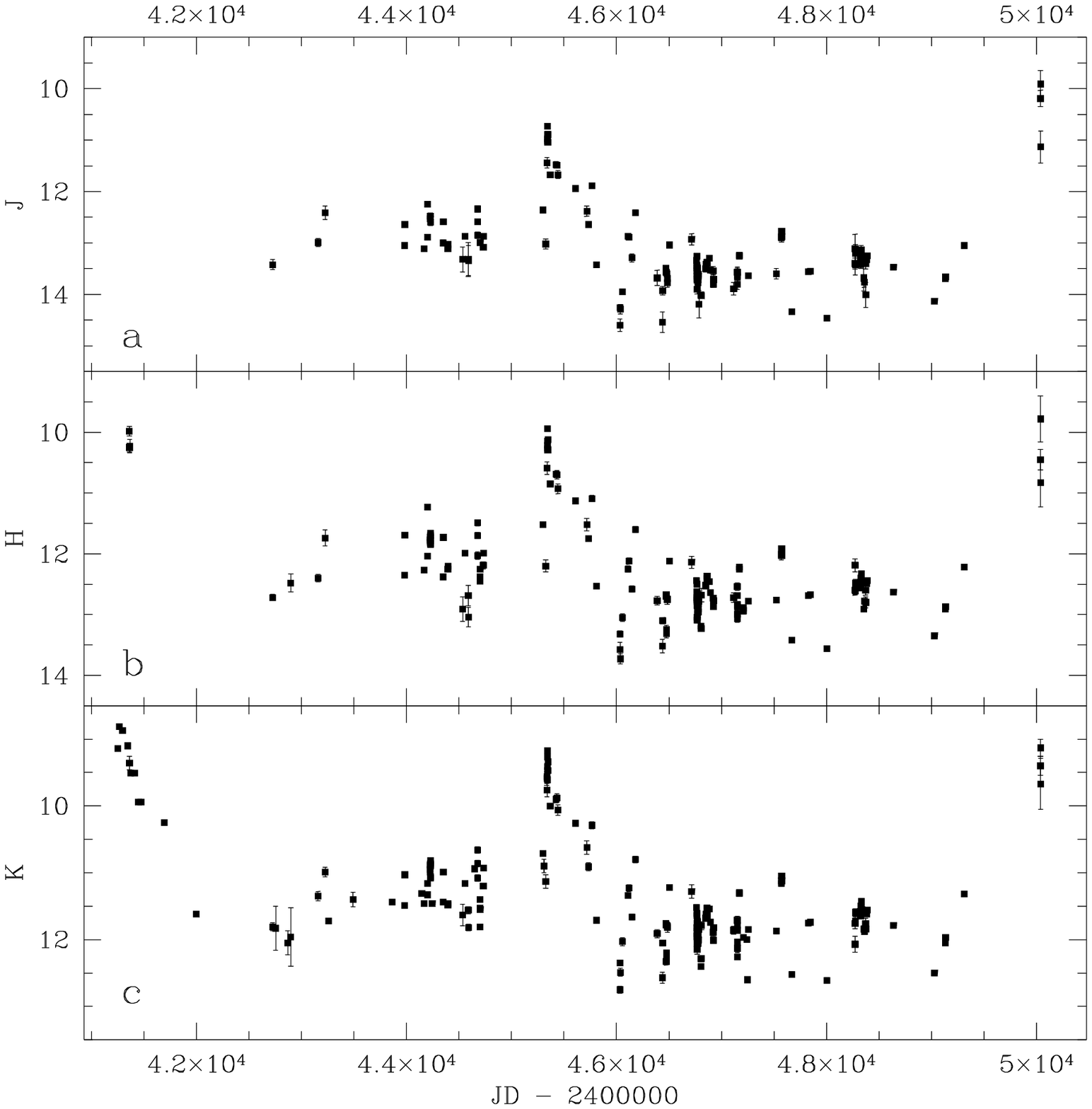}
$$

\caption[]{a: The long-term J  light curve of OJ287 in  covering a period 
of 1975 through 1995, the discontinuity of the light curve 
is due to the lack of observation in this band \\
b: The long-term H light curve of OJ287 covering a period of 1972 through 1995, 
the discontinuity of the light curve is due to the lack of observation in this band\\
c: The long-term K light curve of OJ287  covering a period of 1971 through 1995, 
the discontinuity of the light curve is due to the lack of observation in this band}
\end{figure}

\subsection{Variations}

 From the available literature, we found that the variations in the UBVRIJHK bands are:
 $\Delta U = 4^{m}.72 (11^{m}.86 - 16^{m}.58)$;
 $\Delta B = 5^{m}.93 (12^{m}.47 - 18^{m}.40)$; $\Delta V = 5^{m}.18 (12^{m}.22 - 17^{m}.40)$;
 $\Delta R = 4^{m}.45 (12^{m}.25 - 16^{m}.70)$; $\Delta I = 4^{m}.07 (11^{m}.73 - 15^{m}.80)$;
 $\Delta J = 3^{m}.87 (10^{m}.73 - 14^{m}.60)$; $\Delta H = 3^{m}.78 ( 9^{m}.94 - 13^{m}.73)$;
 $\Delta K = 3^{m}.54 ( 8^{m}.81 - 12^{m}.75)$. 
 The color indices are respectively:  $U-B = -0.60 \pm 0.14 $ ( $N = 391$ pairs ); 
 $B-V =  0.46 \pm 0.17 $ ( $N = 470$ pairs ); 
 $B-I = 1.42  \pm 0.25  $ ( $N = 186$ pairs ); $V-R =  0.48 \pm 0.14  $ ( $N = 243$ pairs ); 
 $V-I =  0.98 \pm 0.22  $ ( $N = 235$ pairs ); $R-I =  0.56 \pm 0.12  $ ( $N = 198 $ pairs ); 
 $J-H =  0.82 \pm 0.16  $ ( $N = 173$ pairs ); $H-K =  0.88 \pm 0.14  $ ( $N = 193$ pairs );
  $J-K =  1.70 \pm 0.19  $ ( $N = 173 $ pairs ).

 From our observations presented in table 1, a correlation has been found
 for B-V and V (see Figure 3):  $V= - (1.02\pm0.08)(B-V)+(15.61\pm 0.01)$ 
 with a correlation coefficient $r= -0.831$. This correlation means that the 
 spectrum flattens when the source brightens. 

\begin{figure}
\epsfxsize=12cm
$$
\epsfbox{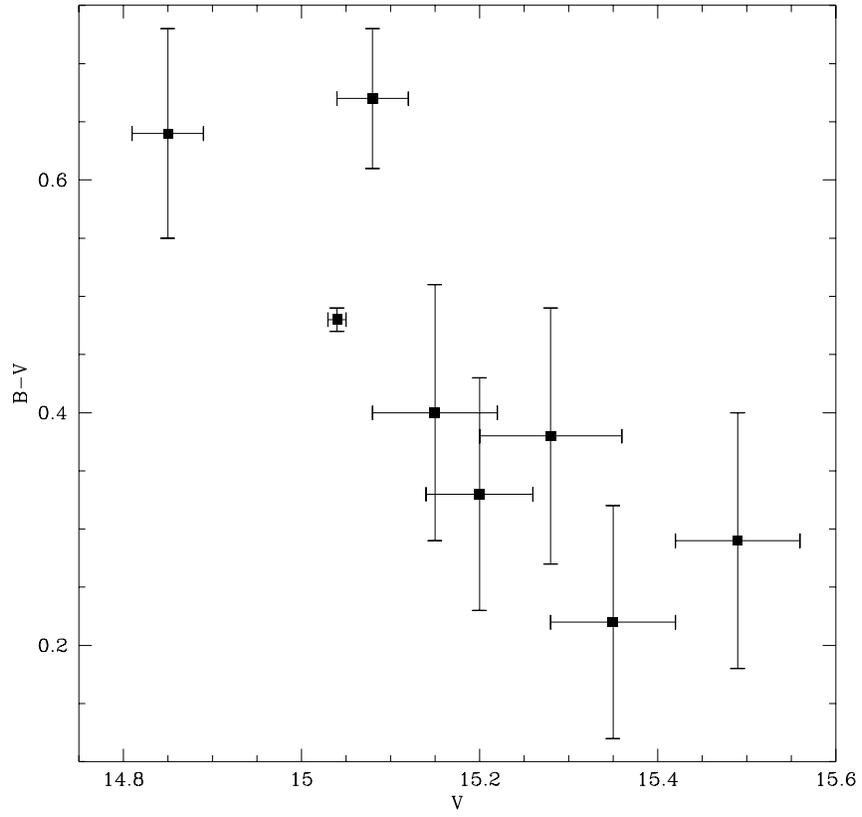}
$$

\caption[]{ Color index--(B-V) versus against V magnitude. 
The relation suggests that the spectrum changes with 
the brightness of the source}
\end{figure}

\section{Correlation Analysis of Variations}
\subsection{DCF method}

 OJ287 is violently variable in the optical and infrared bands. We want to investigate 
 whether the variations in these two bands are correlated or not.  To do so, we use the 
 DCF method described by Edelson \& Krolik (1988) and generalized by Hufnagel \& Bregman (1992)
 which is suitable for data sets which are not sampled evenly or at the same density.

 Following Edelson \& Krolik (1988) and Hufnagel\& Bregman (1992)( see also Tornikoski 
 et al. 1994), we have calculated the DCF.  Firstly, we have calculated the set of 
 unbinned correlation (UDCF) between data points in the two data streams $a$ and $b$

\begin{equation}
 {UDCF_{ij}}={\frac{ (a_{i}- \bar{a}) \times (b_{j}- \bar{b})}{\sqrt{\sigma_{a}^2 \times \sigma_{b}^2}}}
\end{equation}
 where $a_{i}$ and $ b_{j}$ are points in the data sets, $\bar{a}$ and $\bar{b}$ are 
 the means in the data sets, and $\sigma_{a}$ and $\sigma_{b}$ are the standard deviations 
 of each data sets; Secondly, we have averaged the points sharing the same time lag by 
 binning the $UDCF_{ij}$ in suitably sized time-bins to get the $DCF$ for each time lag $\tau$

\begin{equation}
 {DCF(\tau)}=\frac{1}{M}\Sigma UDCF_{ij}(\tau)
\end{equation}
 where $M$ is the number of pairs in the bin.  The standard error for each bin is

\begin{equation}
\sigma (\tau) =\frac{1}{M-1} \{ \Sigma [ UDCF_{ij}-DCF(\tau) ]^{2} \}^{0.5}
\end{equation}

\subsection{Correlation between Optical and Infrared Bands}

 From the light curve, we know that there are more observations in the K band for 
 the infrared data and in the V band for the optical data. So, we use the V and K bands 
 to investigate the correlation between optical and infrared emissions.
 The results are shown in Figure 4 ( a: for bin=30 days and b for bin=50days ). 
 It is clear that the peak in DCF corresponds to ``zero'' point suggesting that 
 the two bands are correlated with almost no time delay.

\begin{figure}
\epsfxsize=15cm
$$
\epsfbox{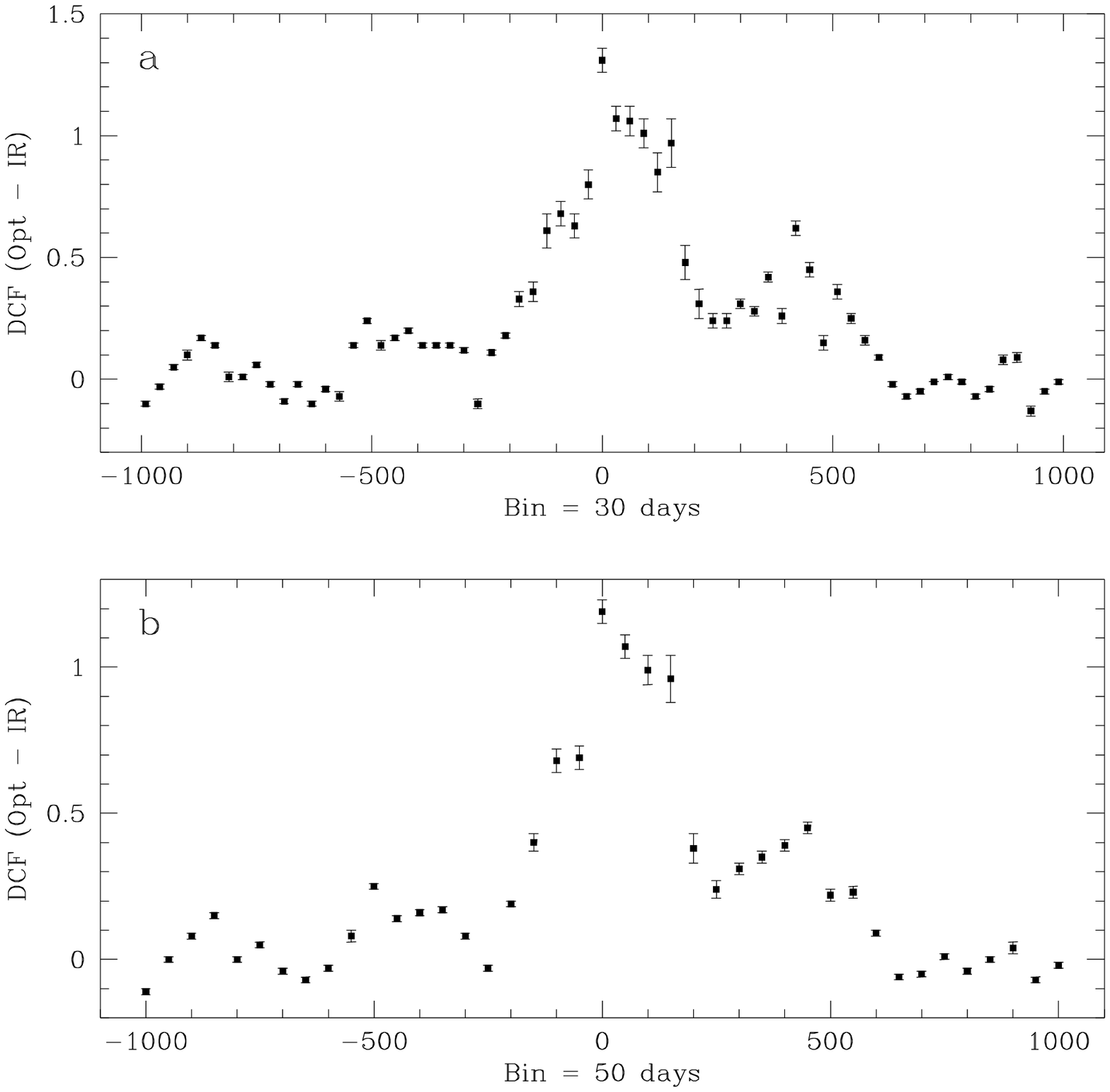}
$$

\caption[]{a: DCF between B and K bands with bin = 30 days\\
b: DCF between B and K bands with bin = 50 days }
\end{figure}

\section{Discussion}

 OJ287 is one of the most extensively studied objects.  Its optical light curve covers 
 a period starting at the end of last century, and shows obviously the outbursts with a 
 period of 12 years, which has been explained by the double black hole model (Sillanpaa 
 et al 1988a).  The detail detection from the OJ-94 project also show clearly the double
 peak structure of the outburst, tentatively explained by Lehto \& Valtonen (1996) and  
 Sillanpaa et al. (1996a).  Besides, the observations from the OJ-94 project has aroused
 much interest in the light curve explanation (Lehto, 1996; Sadun 1996; Valtonen et al. 
 1996; Sundelius et al. 1996; Sillanpaa et al. 1996a). During the project  OJ287 was 
 detected for the first time by EGRET (Pian et al. 1996; Webb et al. 1996), but its 
 radio emission was in a low state during the first optical burst (Valtaoja et al. 1996).

\subsection{Spectral Index}
 OJ287 is violently variable in spectral indices as well as in the flux in the optical 
 and infrared bands.  The long-term analysis of this object showed that the spectra 
 steepened  after 1971 in the optical bands (Takalo \& Sillanpaa 1989) and the infrared 
 spectra steepened during the 1975 to 1990 period ( Zhang \& Xie 1996).  Its spectral 
 indices changed with the brightness of the source.  Takalo \& Sillanpaa (1989) found a 
 strong association between B-V and V magnitude using the available optical data,
 but not for U-B and V. In the present paper, we did not find correlation for U-B and V either. 
 There is no correlation for B-V and U-B, or V-I and I.

 For infrared data, Gear et al (1986a) and Kidger et al. (1994) found a correlation between 
 J-K and J based on the limited data, this tendency has also been reported by Zhang \& Xie 
 (1996).  But during the OJ-94 project, the optical spectral index kept unchanged even during 
 the outburst period (Sillanpaa et al. 1996a), the color indices also stayed constant during the 
 observations of Arimoto et al (1997). It maybe that the correlation between the brightness 
 and the spectral index does not hold during bursts.  From our limited observation, there is a 
 close correlation between  B-V and V ( see Figure 3 ). From the available data, B-V and B-I 
 show the indications of a decrease of spectral index with the time after the 1972  burst; 
 color indices show that the spectrum flattened during the 1983/1984 outbursts.
 But it is interesting to note  that there is also clear spectral flattening during the 
 time of about JD 2448000 when there was no large outburst (see Figure 5). 

\begin{figure}

\epsfxsize=15cm
$$
\epsfbox{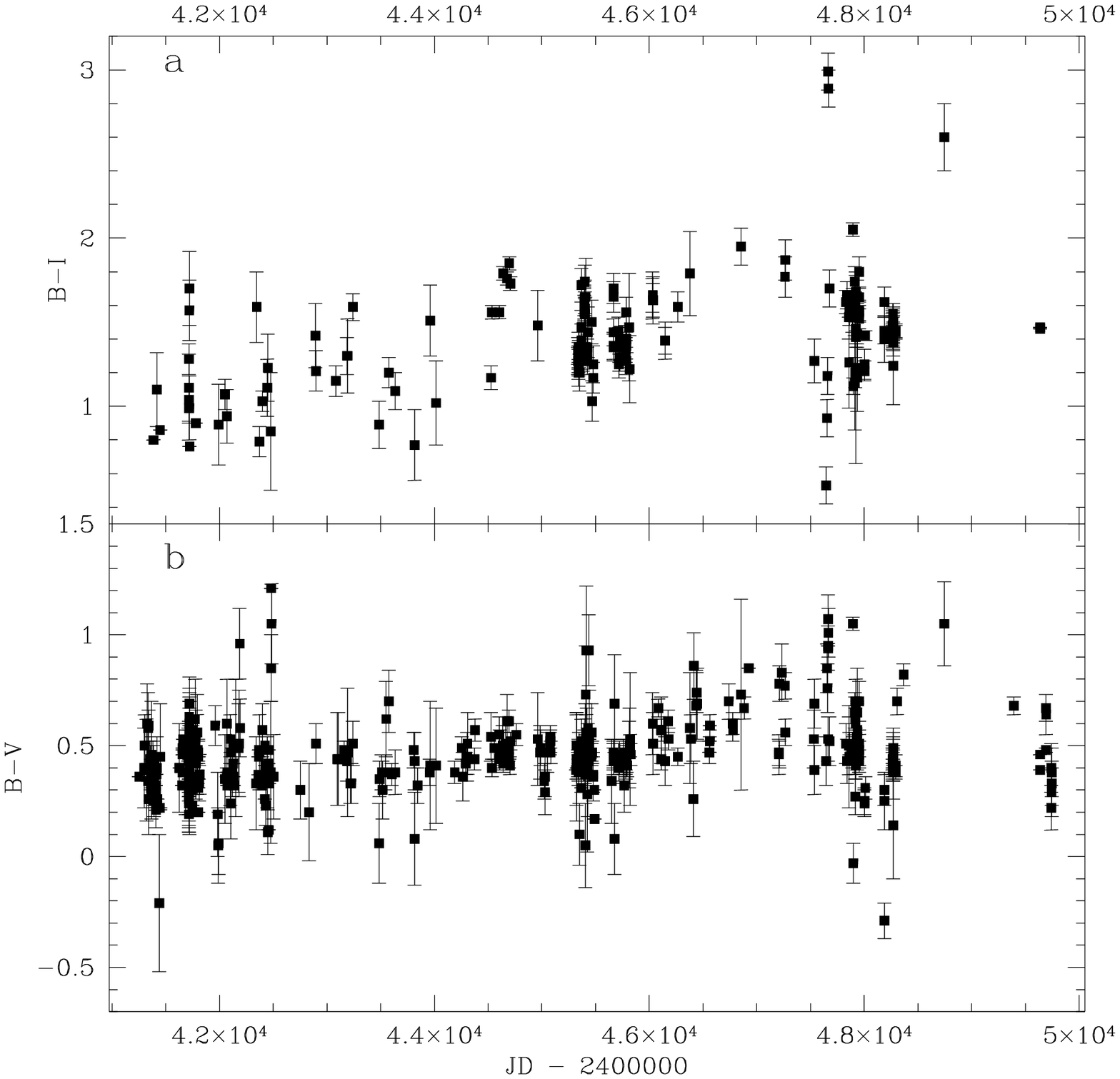}
$$

\caption[]{a. Evolution of Color Index of B-V \\
b. Evolution of Color Index of B-I}
\end{figure}

\subsection{Comparison of Observations}
 In Autumn 1994, Fioruci \& Tosti (1994a,b) announced that OJ287 was at its high level. 
 Arimoto et al. (1997) also obtained its optical data in the V, R and I bands covering a 
 period of October 13, 1994 through May 25, 1996. They obtained that the object was at its
 high level of $V=14^m.03$ in the first half of November 1994 and fainted out to $V=16^m.5$ 
 within 150 days. Comparing our data with theirs, we can see clearly that our data of 
 $V=14^m.85 \pm 0.09$ on Dec. 6 of 1994 is in agreement of their result of 
 $V=14^m.84 \pm 0.03$ on Dec 6.79 of 1994, and our data of $V=15^{m}.32 \pm 0.05$ on 
 January 27 of 1995 is also in agreement with their result of $V=15^m.50 \pm 0.04 $ on 
 January 26.72 of 1995.  For other data we can not compare with theirs, because they got 
 one point every week.  During our observation, V  changed from $15^m.62$ ( Feb. 8, 1994) 
 to $14^m.85$ (Dec 6, 1995), which is also in good agreement with the results obtained by 
 Sillanpaa et al. (1996a), who showed that the range of V was from $14^m.0$ to $16^m.5$.  
 But there are no observation reported in the B band in other literature for the outburst.  
 For our observations, the B magnitudes changed from $ 16^m.07$ (Feb. 7, 1994) to $15^m.49$ 
 (Jan. 01, 1995).  From Figure 1 it is clear that the 1994 optical outburst was fainter than 
 the previous two ( 1972, 1983). But in infrared bands, the peaks for the  three (1972, 1983, 1995)
 outbursts are comparable. Comparing the peaks in the light curve of OJ287 in the paper of  
 Sillanpaa et al. (1996a), we can see that the peak in the 1910s is comparable with the peaks 
 in the 1980s, and the peak in the 1920s is comparable with the peaks in the 1990s.  So, we 
 might have missed a large peak in the 1900s which should be comparable with the peaks of the 
 1970s. If this is true, then there should be a slow variation over about 70 years and we 
 would expect that the following outburst (in 2006) should be brighter than the 1990s outburst.

\subsection{Infrared Observations}
 There is no report in the infrared bands from the object during the OJ-94 project. 
 We had the opportunity to observe it with the 1.26-m infrared telescope during the 
 middle of November 1995. The data shown in table 2 indicate that the infrared emission of the source 
 was at a high level with $K=9^m.13 \pm 0^m.13$ on Nov. 17, 1995. It is comparable
 with the previous observation obtained during the optical outbursts: $K=8^m.8$ in 1972 
 (Soifer \& Neugebauer 1980) and  $K=9^m.25 \pm 0^m.03$ in 1983 (Holmes et al. 1984a ).
 From the three peaks, we can get that the intervals between the successive peaks is about 
 12.0 years, which is consistent with the period derived from the  optical light curves 
 (Sillanpaa et al. 1988a). From the figure in the paper of Sillanpaa et al (1996a) it is clear 
 that our infrared observations correspond to the second peak during the outburst of OJ287.
 So, there should be a missed peak which occurred in the end of 1994!

\subsection{Variability Correlation}

 In order to investigate the emission mechanism from OJ287, variations in the radio, 
 optical and infrared bands have been discussed. Kinman et al(1974) reported that there 
 are some indications of a radio flux increase after the optical and infrared outburst, 
 O'Dell et al. (1978) also reported that the $3mm$ flux  was related with that in the 
 optical and infrared. But it is strange that no corresponding outburst was observed 
 in radio band during OJ-94 project optical outburst. When it was at high level in 
 optical band its radio emission was in a low state (Valtaoja et al. 1996). For the 
 optical and infrared bands, a good correlation was obtained between the
 optical flux and that at 10 $\mu m$ (Rieke \& Kinman 1974) based on the limited data 
 over limited period.  The results shown in this paper indicate that the two bands 
 are strongly correlated with almost no time delay. The reason for that the peak 
 in the DCF is not so sharp is from the fact that the infrared data are fewer than 
 the optical ones and that the clear double-peak in the optical is missed 
 in the infrared observations.

\section{Conclusion}  

 In the present paper, the post-1971 infrared and optical data have been compiled and 
 dealt with by means of the DCF method, the result indicates a strong correlation 
 between these two bands, which suggests that the emission mechanisms in the optical 
 and infrared bands are the same. The color index and brightness association is found 
 from our new optical data.  The time-dependent decreasing color indices post large 
 outburst are also presented in the paper.

\begin{acknowledgements} The authors thank the referee Dr. L. Takalo's comments. 
 This work is supported by the National Scientific Foundation of China (
 the  ninth five-year important project) and the National Pandeng Project of China.
\end{acknowledgements}

\end{document}